\documentclass[
superscriptaddress,
preprint,
 amsmath,amssymb,
 aps,
]{revtex4-1}

\usepackage{graphicx}
\usepackage{dcolumn}
\usepackage{bm}
\usepackage{soul,xcolor}
\usepackage{CJK}

\setstcolor{blue}

\begin{document}
\begin{CJK*}{UTF8}{}

\title{Expanding proton dripline by employing a number of muons}

 \author{Lang Liu {\CJKfamily{gbsn}(刘朗)}}
 \email{liulang@jiangnan.edu.cn}
 \affiliation{School of Science, Jiangnan University, Wuxi 214122, China.}
\author{Yongle Yu {\CJKfamily{gbsn}(余永乐)}}
 \email{yongle.yu@wipm.ac.cn}
 \affiliation{ State Key Laboratory of Magnetic Resonance and Atomic and 
Molecular Physics,\\
  Wuhan Institute of Physics and Mathematics, Chinese Academy of Science,
 West No. 30 Xiao Hong Shan, Wuchang, Wuhan, 430071, China}

\begin{abstract}

Through mean-field calculations, we demonstrate that, 
in a large $Z$ nucleus binding multiple muons, 
these heavy leptons localize within a few dozen
 femtometers of the nucleus. 
 The mutual Coulomb interactions between 
the muons and the  protons can 
lead to a substantial decrease in proton chemical 
potential, surpassing 1 MeV.  
These findings allow for expanding the 
proton-dripline on the nuclear chart in principle, 
suggesting the possible production of nuclei 
with $Z$ around 120.

 
\end{abstract}

\maketitle
\end{CJK*}

\section{Introduction}

Interestingly, the dimensionless value of the electromagnetic 
interaction strength, given by $\frac{e^2}{\hbar c}$, is approximately 
the reciprocal of 137, while the highest observed elemental number 
is 118. The proximity of these fundamental numbers is not coincidental 
but arises from the interplay between nuclear and Coulomb 
interactions in nuclei.
While the short-ranged nuclear interaction strength has a 
dimensionless value of 1, the long-ranged Coulomb repulsive 
interaction between protons, with a strength of 1/137, can 
collectively overwhelm the nuclear interaction and make it 
impossible to bind more protons to a nucleus with $Z$ close to 120.
A rough estimate can be made regarding the energy per proton. 
The Coulomb energy between two unit charges separated by one fm is 
1.44~MeV. However, for a large nuclear system with $Z \approx 100$, 
the average distance between two protons is approximately 8 fm, and 
the Coulomb energy is 0.18~MeV. With 100 protons, the Coulomb energy 
per proton becomes 0.18~MeV *100/2 = 9~MeV (accounting for 
double counting), which corresponds to a typical nuclear binding energy 
per nucleon. This indicates that the chemical potential of protons is 
nearing a sign reversal, and the system is approaching the limit of the largest $Z$.

In atoms, the Coulomb interaction between protons and electrons 
is attractive and may mitigate the repulsive Coulomb energy among 
protons. However, the Coulomb energy contribution from electrons 
is negligible because their average distance from the nucleus is 
several orders of magnitude greater than the typical nuclear distance.
The pronounced discrepancy in size between atoms and nuclei is 
intrinsically tied to the fact that the mass of an electron is 
only about one two-thousandth of the mass of a nucleon. If electrons 
were to be confined within dimensions comparable to the size of a 
nucleus, their kinetic energies would be compelled to 
escalate to the scale of 100~MeV or even higher, 
given their 
diminutive mass. The introduction of muons, as heavier variants of 
electrons in nature, could potentially reshape the scenario 
when coupled with a nucleus. Muons, possessing a mass that 
is one-ninth that of nucleons, can be positioned much closer 
to protons, leading to the emergence of mutual attractive 
Coulomb energies that approach energy scales characteristic 
of nuclear interactions. This rather intriguing possibility could expand 
the scope of nuclear stability studies and enhance our understanding 
of the fundamental forces governing these systems.

The system of a nucleus bound to a muon has been extensively 
studied in the past. It is well established that a muon does 
not participate in strong interactions and interacts with other 
particles through its charge, magnetic moment, and weak and neutral 
currents~\cite{Gorringe2015PPNP84:73--123}. When a muon enters a 
substance, it is slowed down by collisions and is captured by 
an atom, forming a muonic atom. By studying the hyperfine structure 
of the spectrum of a muonic atom, knowledge about the nucleus 
can be obtained, such as determining the nuclear ground state 
spin and measuring the magnetic dipole moment and electric 
quadrupole moment of the nucleus~\cite{Wu1969ARNS19:527--606, 
Knecht2020EPJP135:777, antognini2020measurement, measday2001nuclear}. 
Much theoretical work considering nuclear physics, atomic physics, 
and quantum electrodynamics has also been developed and used to 
study muonic atoms and ions~\cite{Borie1982RMP54:67--118, Dong2011PLB704:600--603, 
acharya2021dispersive, hernandez2019probing}.

In this paper, Skyrme-Hartree-Fock (SHF) is employed to study nuclei 
with a number of muons. SHF is a highly successful self-consistent 
microscopic model extensively used to study nucleus properties~\cite{Vautherin1969PLB29:203--206, 
Vautherin1972PRC5:626--647, Bender2003RMP75:121--180}. Since muons do not participate 
in strong interactions, we only need to consider Coulomb interactions between muons 
and protons, as well as between muons when multiple muons exist.
We demonstrate that when a large-$Z$ nucleus is bound to a number of muons, 
the chemical potential of protons can be lowered by more than 1~MeV, 
indicating that the system can accommodate more protons. This allows 
for the expansion of the proton drip line on the nuclear chart 
and the production of a nucleus with a $Z$ of around 120.
Given that a muon has a lifetime of 2 microseconds, it is 
technically very challenging to generate a nucleus with a 
number of muons in the lab. However, compared to the typical 
nuclear timescale of $10^{-22}$ seconds, such a lifetime is still 
sufficiently long, and these intriguing systems may be 
experimentally investigated in the distant future.

\section{Theoretical Framework}
\label{sec:theory}
We consider a system comprising a nucleus with a specific 
number of muons.  The total Hamiltonian of the nucleus-muons system can be
expressed as:

\begin{equation}\label{eq:totH}
H_{N\mu s}= H_{N} + H_{\mu s}  - 
\int \frac{e^2\rho_p(\mathbf{r})\rho_\mu(\mathbf{r'})}{|\mathbf{r} - \mathbf{r'}|}
  d \mathbf{r}d \mathbf{r'},
\end{equation}

Where $H_{N}$ represents the nuclear Hamiltonian, and 
$H_{\mu s}$ denotes the Hamiltonian of the muons alone; 
$\rho_p(\mathbf{r})$ and $\rho_\mu(\mathbf{r})$ correspond 
to the proton density and muon density, respectively; 
and $e$ represents the charge of a proton.

$H_{\mu s}$ can be written as,
\begin{equation}\label{eq:muon}
H_{\mu s}=  -\sum_{i=1}^{N_\mu} \frac{\hbar^2}{2m_\mu} \nabla^2_i + \sum_{i<j}^{N_\mu}
 \frac{e^2}{|\mathbf{r}_i - \mathbf{r}_j|} 
\end{equation}
where $\hbar$ is the Plank constant, $N_\mu$ corresponds the total number
of muons, and $m_\mu$ denotes the mass of an muon.

The nuclear interaction is modeled as a Skyrme's density-dependent
interaction presented in Ref.~\cite{Vautherin1972PRC5:626--647}.
Here we only give the general framework. The Skyrme interaction
 can be written as a potential
\begin{equation}\label{eq:1}
V=\sum_{i<j} v_{i j}^{(2)}+\sum_{i<j<k} v_{i j k}^{(3)},
\end{equation}
with a two-body part $v_{i j}$ and three-body part $v_{i j k}$.
To simplify calculations, Skyrme used a short-range expansion for 
the two-body interaction and a zero-range force for the three-body force.

For the Skyrme interaction, there exists a straightforward way to obtain 
the Hartree-Fock equations. Consider a nucleus whose ground state is 
represented by a Slater determinant $\phi$ of single-particle states $\phi_i$:
\begin{equation}
\phi\left(x_1, x_2, \ldots, x_A\right)=\frac{1}{\sqrt{A !}} \operatorname{det}\left|\phi_i\left(x_j\right)\right|,
\end{equation}
where $x$ denotes the set $\mathbf{r}, \sigma, q$ of space, 
spin, and isospin coordinates $\left(q=+\frac{1}{2}\right.$ for a proton, $-\frac{1}{2}$ for a neutron). The 
expectation value of the total energy is
\begin{equation}
\begin{aligned}
E= & \langle\phi|(T+V)| \phi\rangle \\
= & \sum_i\left\langle i\left|\frac{p^2}{2 m}\right| i\right\rangle+\frac{1}{2} \sum_{i j}\left\langle i j\left|\tilde{v}_{12}\right| i j\right\rangle  +\frac{1}{6} \sum_{i j k}\left\langle i j k\left|\tilde{v}_{123}\right| i j k\right\rangle \\
= & \int H(\mathbf{r}) \rm d \mathbf{r},
\end{aligned}
\end{equation}
where the notation $\tilde{v}$ denotes an antisymmetrized matrix 
element. For the Skyrme interaction the energy density $H(\mathbf{r})$ is 
an algebraic function of the nucleon densities $\rho_n\left(\rho_p\right)$, the kinetic 
energy $\tau_n\left(\tau_p\right)$, and spin densities $\mathbf{J}_n\left(\mathbf{J}_p\right)$. These 
quantities depend in turn on the single-particle states $\phi_i$ defining 
the Slater-determinant wave function $\phi$,
\begin{equation}
\begin{aligned} & \rho_q(\mathbf{r})=\sum_{i, \sigma}\left|\phi_i(\mathbf{r}, \sigma, q)\right|^2, \\ & \tau_q(\mathbf{r})=
\sum_{i, \sigma}\left|\bm{\nabla} \phi_i(\mathbf{r}, \sigma, q)\right|^2, \\
 & \mathbf{J}_q(\mathbf{r})=(-i) \sum_{i, \sigma, \sigma^{\prime}} \phi_i^*(\mathbf{r}, \sigma, q)\left[\bm{\nabla} \phi_i\left(\mathbf{r}, \sigma^{\prime}, q\right) \times\left\langle\sigma | \bm{\sigma}| \sigma^{\prime}\right\rangle\right] .
\end{aligned}
\end{equation}
The sums in above equations are taken over all occupied single-particle states.
The exact expression for $H(\mathbf{r})$ is following ~\cite{Vautherin1972PRC5:626--647},
\begin{equation}
\begin{aligned}  H(\mathbf{r})= &\frac{\hbar}{2m}\tau(\mathbf{r})+ \frac{1}{2}t_0[ (1+\frac{1}{2}x_0)
\rho^2- (x_0 +\frac{1}{2})(\rho_n^2 + \rho_p^2)] + \frac{1}{4} (t_1+t_2)\rho\tau +
 \\&\frac{1}{8}(t_2-t_1)(\rho_n \tau_n + \rho_p \tau_p) 
+ \frac{1}{16}(t_2- 3t_1)\rho \nabla^2 \rho +  \frac{1}{32}(3t_1 + t_2)(\rho_n \nabla^2 \rho_n + \rho_p \nabla^2 \rho_p)
+ \\& \frac{1}{16}(t_1- t_2)(\mathbf{J}_n^2 + \mathbf{J}_p^2)+ \frac{1}{4}t_3 \rho_n \rho_p \rho + H_C(\mathbf{r})-
  \frac{1}{2}w_0(\rho \bm{\nabla} \cdot\mathbf{J} + \rho_n \bm{\nabla}\cdot\mathbf{J}_n  + \rho_p \bm{\nabla} \cdot\mathbf{J}_p),
\end{aligned}
\end{equation}
where $\rho= \rho_n +\rho_p$, $\tau= \tau_n+ \tau_p$ and $\mathbf{J}=\mathbf{J}_p+\mathbf{J}_p$;  $x_0,t_0,t_1,t_2,t_3,w_0$
 describe the parameterization of the nuclear force. The direct part of Coulomb interaction in
$H_C(\mathbf{r})$ is $\frac{1}{2} V_C(\mathbf{r}) \rho_p(\mathbf{r})$, where
\begin{equation}
V_C(\mathbf{r})=\int \rho_p(\mathbf{r}^{\prime}) \frac{e^2}{\left|\mathbf{r}-\mathbf{r}^{\prime}\right|} \rm d\mathbf{r}^{\prime}.
\end{equation}   
We refer to $V_C(\mathbf{r})$ as the Coulomb potential generated by protons, and one obtains the Coulomb potential of muons by replacing $\rho_{p}$ with  $\rho_{\mu}$.
The Hartree-Fock equations for Skyrme's interaction are obtained by 
writing that the total energy $E$ is stationary with respect to 
individual variations of the single-particle states $\phi_i$, with 
the subsidiary condition that $\phi_i$ are normalized
\begin{equation}
\frac{\delta}{\delta \phi_i}\left(E-\sum_i e_i \int\left|\phi_i(\mathbf{r})\right|^2 \rm d^3 r\right)=0 .
\end{equation}
It can be shown $\phi_i$ statify the following set of equations,
 \begin{equation}
 [ -\bm{\nabla} \cdot \frac{\hbar^2}{2 m^*_q(\mathbf{r})}\bm{\nabla} + U_q(\mathbf{r})
+  \mathbf{W}_q(\mathbf{r})\cdot (-i)(\bm{\nabla} \times  \bm{\sigma})] \phi_i = e_i \phi_i.
\label{localSchro}
\end{equation}
Eq.\ref{localSchro} involves an effective mass $m^*_q (\mathbf{r})$
which depends on the density,
\begin{equation}
\frac{\hbar^2}{2 m^*_q (\mathbf{r})}= \frac{\hbar^2}{2 m_q}+ \frac{1}{4}(t_1 +t_2)\rho + \frac{1}{8}
(t_2 - t_1) \rho_q.
\end{equation}
The potential $U_q(\mathbf{r})$ is expressed as
following,
\begin{equation}
\begin{aligned}
U_q(\mathbf{r})=  & t_0[ (1 +\frac{1}{2}x_0)\rho- (x_0 +\frac{1}{2})\rho_q] + 
 \frac{1}{4}t_3 (\rho^2 - \rho_q^2) - \frac{1}{8} (3t_1 - t_2) \nabla^2 \rho + \frac{1}{16} (3t_1 + t_2) \nabla^2 \rho_q
 +\\ & \frac{1}{4} (t_1 + t_2) \tau +  \frac{1}{8} (t_2 - t_1) \tau_q - \frac{1}{2} W_0( \bm{\nabla}\cdot \mathbf{J}
+ \bm{\nabla}\cdot \mathbf{J}_q) + \delta_{q,+\frac{1}{2}}V_C(\mathbf{r}).
\end{aligned}
 \end{equation}
The form factor $\mathbf{W}_q(\mathbf{r})$ of the spin-orbit potential is 
\begin{equation}
\mathbf{W}_q(\mathbf{r})= \frac{1}{2}W_0 (\bm{\nabla}\rho + \bm{\nabla}\rho_q)+ 
\frac{1}{8}(t_1- t_2)\mathbf{J}_q(\mathbf{r}).
\end{equation}

We employ the force II parameterization from Ref.~\cite{Vautherin1972PRC5:626--647} 
for the Skyrme force in the numerical code. Specifically, we use the following parameter 
values: $x_0= 0.34$, $t_0$= -1169.9~MeV fm$^3$, $t_1$= 585.6~MeV fm$^5$,
$t_2$= -27.1~MeV fm$^5$, $t_3$ = 9331.1~MeV fm$^6$,  and $W_0$= 105~MeV fm$^5$.  
 By successfully reproducing the outcomes 
reported in the reference for various nuclei, we validate the reliability of the code.

When binding a specific number of muons to the nuclei, it 
becomes imperative to incorporate the Coulomb potential 
contributed by the muons into the self-consistent mean field
 calculation for determining the single-particle orbit of 
protons. 
The mean field that governs the single-particle orbits of 
muons comprises the Coulomb potential generated by the 
protons and the Coulomb potential generated by the muons themselves,
specifically, single-particle states $\varphi_i(\mathbf{r},\sigma)$
satisfies the following equations,
\begin{equation}
( -\frac{\hbar^2}{2 m_\mu} {\nabla}^2  - \int \frac{e^2(\rho_p(\mathbf{r'})-\rho_\mu(\mathbf{r'}))}{|\mathbf{r} - \mathbf{r'}|}
  d \mathbf{r'}) \varphi_i(\mathbf{r},\sigma)
 = \varepsilon_i \varphi_i(\mathbf{r},\sigma).
\end{equation}

To obtain self-consistent results, we perform a series 
of numerical iterations until convergence is achieved. 
During each iteration, the updated potentials derived from the
previous iteraction are employed 
to compute the single-particle orbits. From these orbits, 
the new single-particle densities are calculated and utilized 
to construct the updated potentials for the next iteration. 
This iterative process continues until the desired convergence is attained.

In the numerical implementation, we assume the spherical symmetry
of the system so that the computation is reduced essentially to 
integrate  the system
along the radial direction. We use a lattice system to model the  
radial dimension and the lattice constant can be smaller as 0.08 fm
while the radial size of the system can extent up to 60 fm.

Given our objective of numerically estimating the shift in proton 
chemical potentials resulting from the presence of muons, the 
simplifications made in our model can be justified. These 
simplifications 
include the assumption of spherical symmetry, the ignorance of
nuclear pairing interactions and non-relativistic treament
of muons.


\section{Results and Discussion}
\label{sec:resu}

\begin{figure}[ht!]
\centering
 \includegraphics[scale=1]{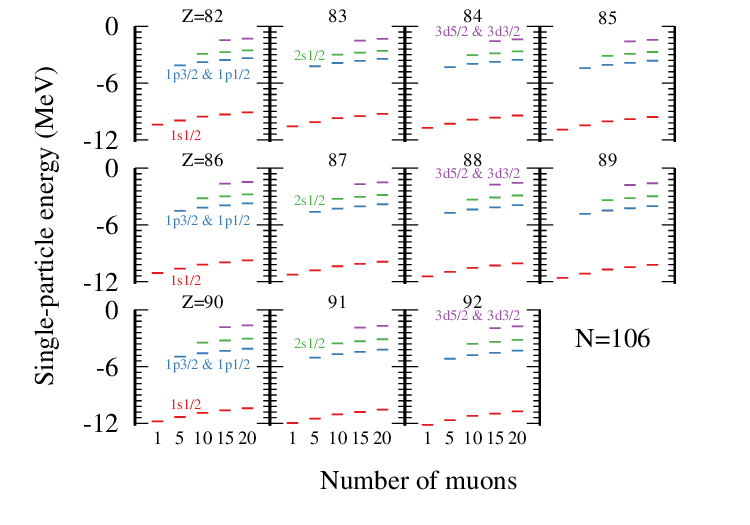}
\caption{(Color online) The single-particle levels of muons as a function of the number of muons for $N$=106 isotones ($Z$ from 82 to 92). The red, blue, green, and purple lines represent $1s_{1/2}$, $1p_{3/2}$ ($1p_{1/2}$), $2s_{1/2}$, and $3d_{5/2}$ ($3d_{3/2}$) orbitals, respectively.}
\label{fig_1}
\end{figure}

As shown in Fig.~\ref{fig_1}, we calculated the single-particle levels of muons for $N$=106 isotones ($Z$ from 82 to 92). The red, blue, green, and purple lines represent $1s_{1/2}$, $1p_{3/2}$ ($1p_{1/2}$), $2s_{1/2}$, and $3d_{5/2}$ ($3d_{3/2}$) orbitals of muons, respectively.
The $1p_{3/2}$ and $1p_{1/2}$ orbitals are degenerate since the deformation is ignored, as are the $3d_{5/2}$ and $3d_{3/2}$ orbitals.
We found that, for each nucleus, the muon single-particle energy level increases with the number of muons since more muons also give stronger repulsion among themselves.
For example, the energy of the $1s_{1/2}$ orbital for $Z = 82$ goes from -10.39~MeV at $N_\mu=1$ to
 -9.10~MeV at $N_\mu= 20$.
For a certain number of muons, for instance, at $N_\mu= 1$, the single-particle energy of $1s_{1/2}$ decreases monotonically with the increase of $Z$, that is, from $-10.39$ MeV at $Z = 82$ to $-12.14$ MeV at $Z = 92$. For another example, at $N_\mu= 10$, the energy level of muons $1s_{1/2}$ decreases gradually from -9.54 MeV at $Z = 82$ to $-11.21$ MeV at $Z = 92$. The energy level of $1p_{3/2}$ decreases from -3.78 MeV to - 4.79 MeV. The $2s_{1/2}$ energy level drops from -2.92 MeV to -3.60 MeV. This decreasing trend in energy levels holds true for other numbers of muons as well. Overall, as the number of protons increases, the single-particle energy levels of muons become slightly more negative. This can be attributed to the fact that more protons exert a stronger attractive force on the muons.
There exists a large energy gap of about 6 MeV 
between the $1s_{1/2}$ orbital and the $1p_{3/2}$ ($1p_{1/2}$) orbital for the whole isotones.


\begin{figure}[ht!]
\centering
 \includegraphics[scale=1]{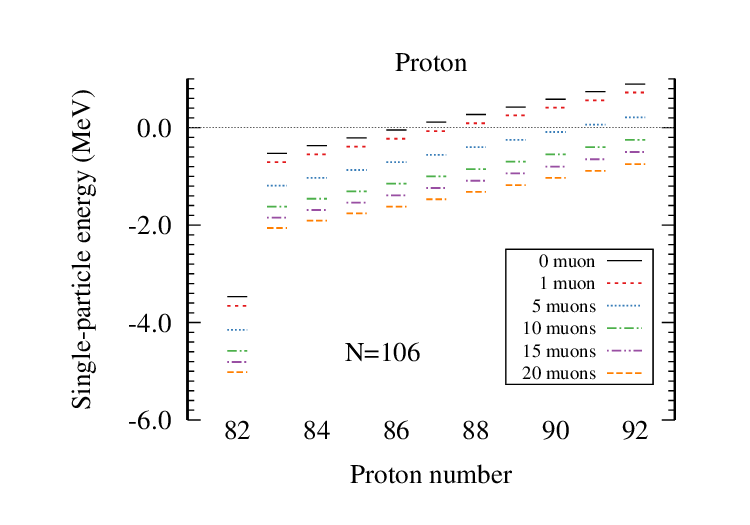}
 \caption{(Color online) The energy of the last single-particle level of protons with the different numbers of muons as a function of the proton number for $N$=106 isotones. The solid black lines, dashed red lines, dotted blue lines, dash-dot green lines, dash-dot-dot purple lines, and long dashed orange lines represent the number of muons 0, 1, 5, 10, 15, and 20, respectively.}
\label{fig_2}
\end{figure}

We have also calculated the single-particle levels of protons with different numbers of muons. Fig.~\ref{fig_2} shows the energy of the last single-particle level of protons as a function of the proton number for $N=106$ isotones. The solid black lines, dashed red lines, dotted blue lines, dash-dot green lines, dash-dot-dot purple lines, and long dashed orange lines represent the numbers of muons 0, 1, 5, 10, 15, and 20, respectively.  There is a clear energy gap between $Z=82$ and 83. Additionally, as the number of protons increases, the energy of the proton level becomes larger. Beyond proton number 87, the energy of the last energy level of the proton is greater than zero (at $N_\mu=0$). What's  interesting is that when muons are considered, the proton energy level can decrease significantly. The magnitude of the decrease enhances with the increase in the number of muons. Our calculations have revealed that adding one muon can reduce the proton energy level by about 0.2 MeV, and 20 muons can reduce the proton energy level by about 1.7~MeV. Therefore, when muons are introduced, for nuclei with $Z=87$, the last unbound energy level of the proton becomes a bound level. More muons mean more bound proton levels. After introducing 10 muons, the last proton energy levels of all $N=106$ isotones become bound single-particle states. Hence, introducing muons in experiments may enable the extrapolation of the proton drip line to obtain more proton-rich nuclei.


\begin{figure}[ht!]
  \centering
    \includegraphics[scale=1]{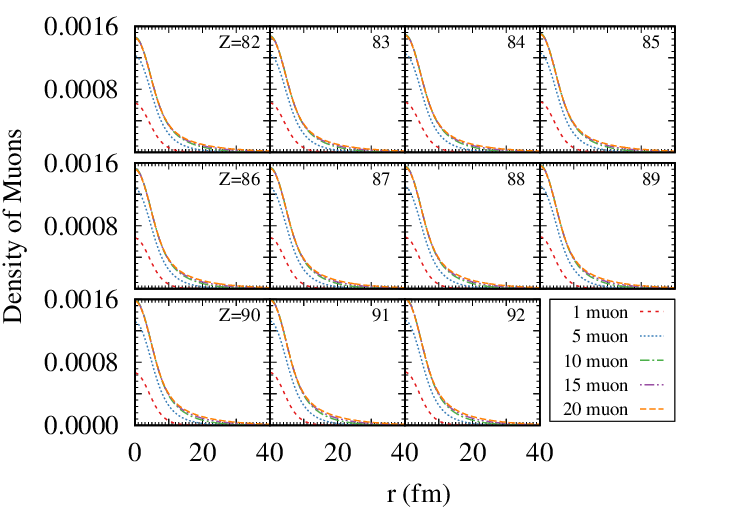}
  \caption{(Color online) The density distribution of muons with the different numbers of muons for $N=106$ isotones. The dashed red lines, dotted blue lines, dash-dot green lines, dash-dot-dot purple lines, and long dashed orange lines represent the number of muons 1, 5, 10, 15, and 20, respectively.}
\label{fig_3}
\end{figure}
The density distribution of muons with the different numbers of muons for $N=106$ isotones has been calculated and shown in Fig.~\ref{fig_3}.
The dashed red lines, dotted blue lines, dash-dot green lines, dash-dot-dot purple lines, and long dashed orange lines represent the number of muons 1, 5, 10, 15, and 20, respectively.
Although the density of muons diffuses to the space beyond 30~fm, the primary density is still distributed within 10~fm. Thus there is considerable overlap with the nucleus.
At the same time, as the number of protons increases, the center density of muons also enlarges by about $10\%$, indicating that muons are attracted closer to the nucleus.


\begin{figure}[ht!]
  \centering
    \includegraphics[scale=1]{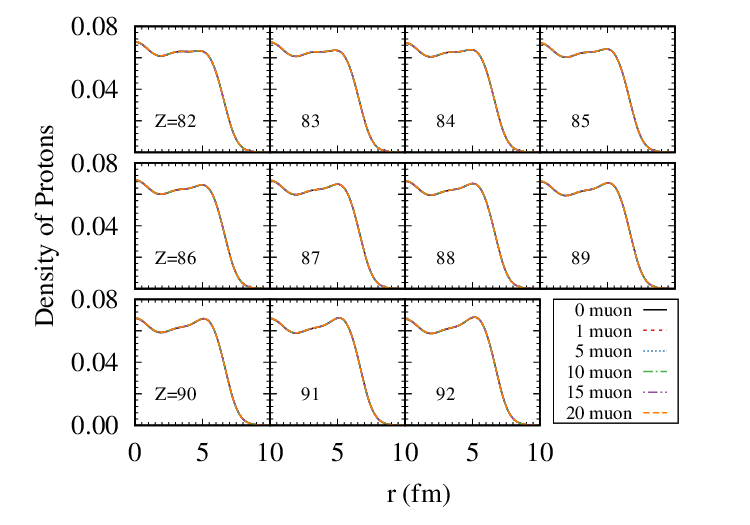}
  \caption{(Color online) The density distribution of protons with the different numbers of muons for $N=106$ isotones. The solid black lines, dashed red lines, dotted blue lines, dash-dot green lines, dash-dot-dot purple lines, and long dashed orange lines represent the number of muons 0, 1, 5, 10, 15, and 20, respectively.}
\label{fig_4}
\end{figure}
Fig.~\ref{fig_4} shows the density distribution of protons with the different numbers of muons for $N=106$ isotones. 
The solid black lines, dashed red lines, dotted blue lines, dash-dot green lines, dash-dot-dot purple lines, and long dashed orange lines represent the number of muons 0, 1, 5, 10, 15, and 20, respectively.
For the whole isotones, the density of protons is mainly distributed between 0 and 8 fm, so there can be a significant overlap with the density of muons. 



\begin{figure}[ht!]
  \centering
    \includegraphics[scale=1]{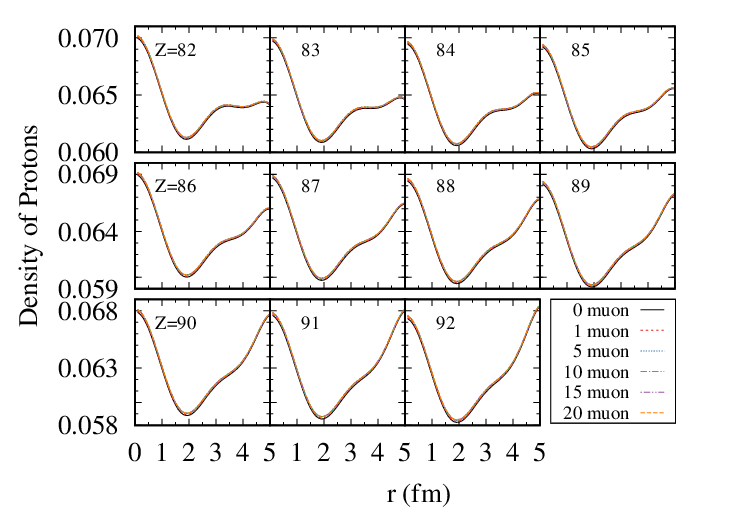}
  \caption{(Color online) The same as Fig.~\ref{fig_4}, but it is the enlarged ones in the coordinate $r$ from 0 to 5~fm.}
\label{fig_5}
\end{figure}
In order to further study the influence of muons on the proton density, we zoomed in on the above plots.
Fig.~\ref{fig_5} is the enlarged ones of Fig.~\ref{fig_4} in the coordinate $r$ from 0 to 5~fm.
As discussed above, the proton center density varies by less than $3\%$ across the isotones. 
The change of proton density due to different number of muons is minor, about $0.3\%$.

\begin{figure}[ht!]
\centering
    \includegraphics[scale=1]{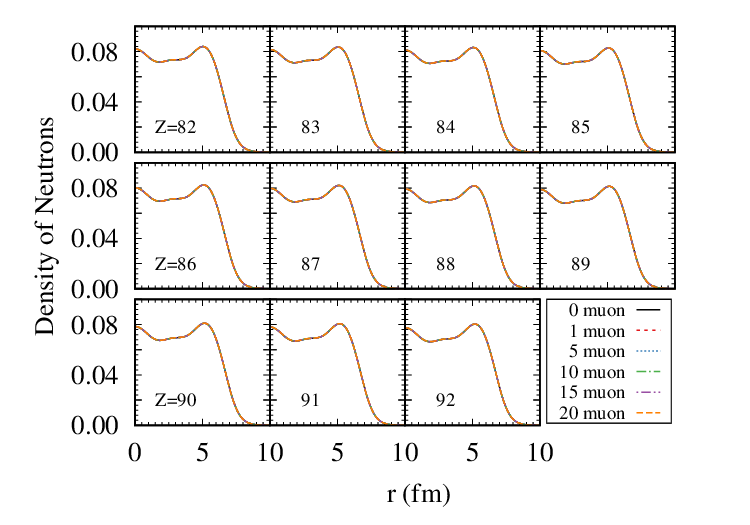}
  \caption{(Color online) The density distribution of neutrons with the different numbers of muons for $N=106$ isotones. The solid black lines, dashed red lines, dotted blue lines, dash-dot green lines, dash-dot-dot purple lines, and long dashed orange lines represent the number of muons 0, 1, 5, 10, 15, and 20, respectively.}
\label{fig_6}
\end{figure}
\begin{figure}[ht!]
\centering
    \includegraphics[scale=1]{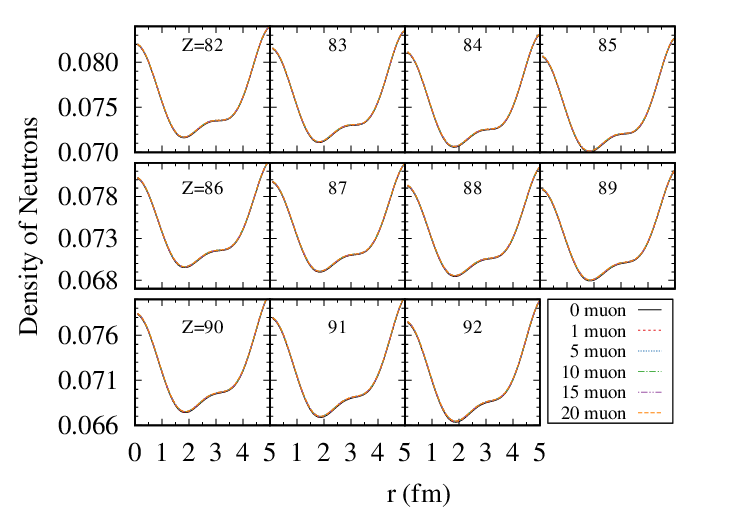}
  \caption{(Color online) The same as Fig.~\ref{fig_6}, but it is the enlarged ones in the coordinate $r$ from 0 to 5 fm.}
\label{fig_7}
\end{figure}

The density distribution of neutrons with the different numbers of muons for $N=106$ isotones is presented in Fig.~\ref{fig_6}.
Fig.~\ref{fig_7} is a partial enlargement of Fig.~\ref{fig_6} from 0 to 5~fm.
The solid black lines, dashed red lines, dotted blue lines, dash-dot green lines, dash-dot-dot purple lines, and long dashed orange lines represent the number of muons 0, 1, 5, 10, 15, and 20, respectively.
The calculated neutron density is also mainly concentrated in $0\sim 8$~fm, which can overlap significantly with the muons density.
However, compared to the case of protons, muons have a much weaker, almost negligible effect on neutron density.


\begin{figure}[ht!]
  \centering
    \includegraphics[scale=1]{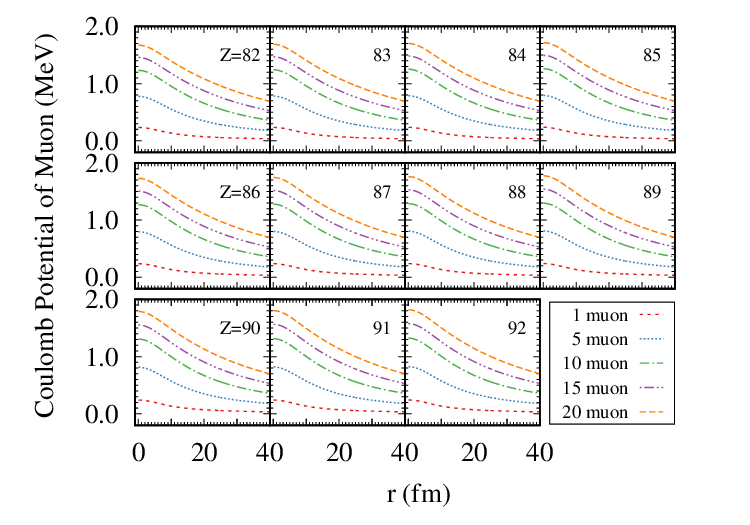}
  \caption{(Color online) The Coulomb potential generated by $\rho_{\mu}$ with the different numbers of muons for $N=106$ isotones. The dashed red lines, dotted blue lines, dash-dot green lines, dash-dot-dot purple lines, and long dashed orange lines represent the number of muons 1, 5, 10, 15, and 20, respectively. Here, the absolute value of the Coulomb potential is taken.}
\label{fig_8}
\end{figure} 
Fig.~\ref{fig_8} shows the Coulomb potential generated by $\rho_{\mu}$ with the different numbers of muons for $N=106$ isotones. 
The dashed red lines, dotted blue lines, dash-dot green lines, dash-dot-dot purple lines, and long dashed orange lines represent the number of muons 1, 5, 10, 15, and 20, respectively (The Coulomb potential is calculated with Eq.~(8) by substituting into the muon density).
Although the muon is negatively charged, here we take its absolute value.
It can be found that the Coulomb potential of the muons increases by about $3.5\%$ as the number of protons increases from 82 to 92. 
However, for a particular nucleus, increasing the number of muons from 1 to 20 increases the Coulomb potential of the muons by about 1.5~MeV.
For example, for a nucleus with $Z=87$, The Coulomb potential of muons near the center, corresponding to the number of muons being 1, 5, 10, 15, and 20, is 0.2, 0.8, 1.3, 1.5, and 1.7 MeV, which is approximately the same as the change of 
energy of the proton Fermi level.


\begin{figure}[ht!]
  \centering
    \includegraphics[scale=1]{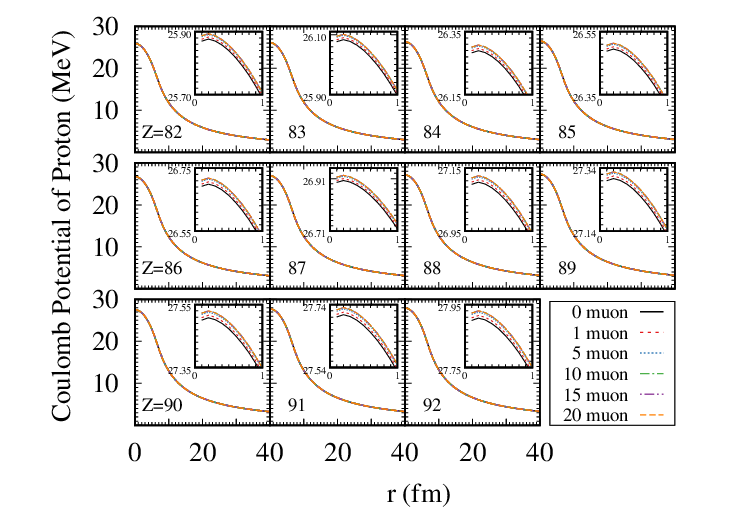}
  \caption{(Color online) The Coulomb potential generated by $\rho_{\rm p}$ with the different numbers of muons for $N=106$ isotones. The solid black lines, dashed red lines, dotted blue lines, dash-dot green lines, dash-dot-dot purple lines, and long dashed orange lines represent the number of muons 0, 1, 5, 10, 15, and 20, respectively. Inserts: the enlarged subfigure in the coordinate $r$ from 0 to 1 fm.}
\label{fig_9}
\end{figure}
The Coulomb potential generated by $\rho_{\rm p}$ with the different numbers of muons for $N=106$ isotones is shown in Fig.~\ref{fig_9}. 
The inserts indicate the enlarged ones in the coordinate $r$ from 0 to 1~fm.
The solid black lines, dashed red lines, dotted blue lines, dash-dot green lines, dash-dot-dot purple lines, and long dashed orange lines represent the number of muons 0, 1, 5, 10, 15, and 20, respectively.
We found that as $Z$ increases from 82 to 92, the Coulomb potential of the protons increases by about $8\%$. 
However, for a particular nucleus, increasing $N_\mu$ from 1 to 20 changes the Coulomb potential of the proton by less than $0.1\%$.


\section{Summary}
\label{sec:sum}
In this work, we studied the properties of the nucleus in the muon atom utilizing a spherical mean-field calculation with Skyrme interaction.
Taking the $N=106$ isotones as an example, we investigate the influence of the muons on the nuclear structure.
It is found that the single-particle levels of muons decrease with the increase in the number of protons and rise with the increase in the number of muons.
More importantly, we found that, although the proton Fermi level changes from bound to unbound with the increase of $Z$, the addition of muons significantly reduces the Fermi level of the proton. 
Moreover, increasing the muons from 0 to 20 even lowers the proton Fermi level by 1.7~MeV.
This could allow for the expansion of the proton dripline on the nuclear chart and the production of a nucleus with a $Z$ of approximately 120.
We analyzed the effect of muons on the proton and neutron density distribution. 
We found that the neutron density is hardly affected by muons, while the proton density changes by approximately 0.3\% in the range of 0 to 5 fm due to the influence of muons. Additionally, the Coulomb potential caused by the protons changes by about 0.1\% due to the influence of muons. However, the Coulomb potential generated by muons can provide energies ranging from 0.2 to 1.7~MeV inside the nucleus, and this order of magnitude is roughly equivalent to the drop in the energy  of the proton Fermi level.

\begin{acknowledgments}
This work was supported by the Natural Science Foundation of China (Grants No. 11775099) 

\end{acknowledgments}

\bibliographystyle{apsrev4-1}
\bibliography{apssamp_interaction}.

\begin{thebibliography}{12}%
\makeatletter
\providecommand \@ifxundefined [1]{%
 \@ifx{#1\undefined}
}%
\providecommand \@ifnum [1]{%
 \ifnum #1\expandafter \@firstoftwo
 \else \expandafter \@secondoftwo
 \fi
}%
\providecommand \@ifx [1]{%
 \ifx #1\expandafter \@firstoftwo
 \else \expandafter \@secondoftwo
 \fi
}%
\providecommand \natexlab [1]{#1}%
\providecommand \enquote  [1]{``#1''}%
\providecommand \bibnamefont  [1]{#1}%
\providecommand \bibfnamefont [1]{#1}%
\providecommand \citenamefont [1]{#1}%
\providecommand \href@noop [0]{\@secondoftwo}%
\providecommand \href [0]{\begingroup \@sanitize@url \@href}%
\providecommand \@href[1]{\@@startlink{#1}\@@href}%
\providecommand \@@href[1]{\endgroup#1\@@endlink}%
\providecommand \@sanitize@url [0]{\catcode `\\12\catcode `\$12\catcode
  `\&12\catcode `\#12\catcode `\^12\catcode `\_12\catcode `\%12\relax}%
\providecommand \@@startlink[1]{}%
\providecommand \@@endlink[0]{}%
\providecommand \url  [0]{\begingroup\@sanitize@url \@url }%
\providecommand \@url [1]{\endgroup\@href {#1}{\urlprefix }}%
\providecommand \urlprefix  [0]{URL }%
\providecommand \Eprint [0]{\href }%
\providecommand \doibase [0]{http://dx.doi.org/}%
\providecommand \selectlanguage [0]{\@gobble}%
\providecommand \bibinfo  [0]{\@secondoftwo}%
\providecommand \bibfield  [0]{\@secondoftwo}%
\providecommand \translation [1]{[#1]}%
\providecommand \BibitemOpen [0]{}%
\providecommand \bibitemStop [0]{}%
\providecommand \bibitemNoStop [0]{.\EOS\space}%
\providecommand \EOS [0]{\spacefactor3000\relax}%
\providecommand \BibitemShut  [1]{\csname bibitem#1\endcsname}%
\let\auto@bib@innerbib\@empty
\bibitem [{\citenamefont {Gorringe}\ and\ \citenamefont
  {Hertzog}(2015)}]{Gorringe2015PPNP84:73--123}%
  \BibitemOpen
  \bibfield  {author} {\bibinfo {author} {\bibfnamefont {T.}~\bibnamefont
  {Gorringe}}\ and\ \bibinfo {author} {\bibfnamefont {D.}~\bibnamefont
  {Hertzog}},\ }\href {\doibase http://dx.doi.org/10.1016/j.ppnp.2015.06.001}
  {\bibfield  {journal} {\bibinfo  {journal} {Prog. Part. Nucl. Phys.}\
  }\textbf {\bibinfo {volume} {84}},\ \bibinfo {pages} {73} (\bibinfo {year}
  {2015})}\BibitemShut {NoStop}%
\bibitem [{\citenamefont {Wu}\ and\ \citenamefont
  {Wilets}(1969)}]{Wu1969ARNS19:527--606}%
  \BibitemOpen
  \bibfield  {author} {\bibinfo {author} {\bibfnamefont {C.~S.}\ \bibnamefont
  {Wu}}\ and\ \bibinfo {author} {\bibfnamefont {L.}~\bibnamefont {Wilets}},\
  }\href {\doibase https://doi.org/10.1146/annurev.ns.19.120169.002523}
  {\bibfield  {journal} {\bibinfo  {journal} {Ann. Rev. Nucl. Sci.}\ }\textbf
  {\bibinfo {volume} {19}},\ \bibinfo {pages} {527} (\bibinfo {year}
  {1969})}\BibitemShut {NoStop}%
\bibitem [{\citenamefont {Knecht}\ \emph {et~al.}(2020)\citenamefont {Knecht},
  \citenamefont {Skawran},\ and\ \citenamefont
  {Vogiatzi}}]{Knecht2020EPJP135:777}%
  \BibitemOpen
  \bibfield  {author} {\bibinfo {author} {\bibfnamefont {A.}~\bibnamefont
  {Knecht}}, \bibinfo {author} {\bibfnamefont {A.}~\bibnamefont {Skawran}}, \
  and\ \bibinfo {author} {\bibfnamefont {S.~M.}\ \bibnamefont {Vogiatzi}},\
  }\href {\doibase https://doi.org/10.1140/epjp/s13360-020-00777-y} {\bibfield
  {journal} {\bibinfo  {journal} {Eur. Phys. J. Plus}\ }\textbf {\bibinfo
  {volume} {135}},\ \bibinfo {pages} {777} (\bibinfo {year}
  {2020})}\BibitemShut {NoStop}%
\bibitem [{\citenamefont {Antognini}\ \emph {et~al.}(2020)\citenamefont
  {Antognini} \emph {et~al.}}]{antognini2020measurement}%
  \BibitemOpen
  \bibfield  {author} {\bibinfo {author} {\bibfnamefont {A.}~\bibnamefont
  {Antognini}} \emph {et~al.},\ }\href {\doibase 10.1103/PhysRevC.101.054313}
  {\bibfield  {journal} {\bibinfo  {journal} {Phys. Rev. C}\ }\textbf {\bibinfo
  {volume} {101}},\ \bibinfo {pages} {054313} (\bibinfo {year}
  {2020})}\BibitemShut {NoStop}%
\bibitem [{\citenamefont {Measday}(2001)}]{measday2001nuclear}%
  \BibitemOpen
  \bibfield  {author} {\bibinfo {author} {\bibfnamefont {D.~F.}\ \bibnamefont
  {Measday}},\ }\href {\doibase 10.1016/S0370-1573(01)00012-6} {\bibfield
  {journal} {\bibinfo  {journal} {Physics Reports}\ }\textbf {\bibinfo {volume}
  {354}},\ \bibinfo {pages} {243} (\bibinfo {year} {2001})}\BibitemShut
  {NoStop}%
\bibitem [{\citenamefont {Borie}\ and\ \citenamefont
  {Rinker}(1982)}]{Borie1982RMP54:67--118}%
  \BibitemOpen
  \bibfield  {author} {\bibinfo {author} {\bibfnamefont {E.}~\bibnamefont
  {Borie}}\ and\ \bibinfo {author} {\bibfnamefont {G.~A.}\ \bibnamefont
  {Rinker}},\ }\href {\doibase 10.1103/RevModPhys.54.67} {\bibfield  {journal}
  {\bibinfo  {journal} {Rev. Mod. Phys.}\ }\textbf {\bibinfo {volume} {54}},\
  \bibinfo {pages} {67} (\bibinfo {year} {1982})}\BibitemShut {NoStop}%
\bibitem [{\citenamefont {Dong}\ \emph {et~al.}(2011)\citenamefont {Dong},
  \citenamefont {Zuo}, \citenamefont {Zhang}, \citenamefont {Scheid},
  \citenamefont {Gu},\ and\ \citenamefont {Wang}}]{Dong2011PLB704:600--603}%
  \BibitemOpen
  \bibfield  {author} {\bibinfo {author} {\bibfnamefont {J.}~\bibnamefont
  {Dong}}, \bibinfo {author} {\bibfnamefont {W.}~\bibnamefont {Zuo}}, \bibinfo
  {author} {\bibfnamefont {H.}~\bibnamefont {Zhang}}, \bibinfo {author}
  {\bibfnamefont {W.}~\bibnamefont {Scheid}}, \bibinfo {author} {\bibfnamefont
  {J.}~\bibnamefont {Gu}}, \ and\ \bibinfo {author} {\bibfnamefont
  {Y.}~\bibnamefont {Wang}},\ }\href {\doibase
  doi:10.1016/j.physletb.2011.09.057} {\bibfield  {journal} {\bibinfo
  {journal} {Phys. Lett. B}\ }\textbf {\bibinfo {volume} {704}},\ \bibinfo
  {pages} {600} (\bibinfo {year} {2011})}\BibitemShut {NoStop}%
\bibitem [{\citenamefont {Acharya}\ \emph {et~al.}(2021)\citenamefont
  {Acharya}, \citenamefont {Lensky}, \citenamefont {Bacca}, \citenamefont
  {Gorchtein},\ and\ \citenamefont {Vanderhaeghen}}]{acharya2021dispersive}%
  \BibitemOpen
  \bibfield  {author} {\bibinfo {author} {\bibfnamefont {B.}~\bibnamefont
  {Acharya}}, \bibinfo {author} {\bibfnamefont {V.}~\bibnamefont {Lensky}},
  \bibinfo {author} {\bibfnamefont {S.}~\bibnamefont {Bacca}}, \bibinfo
  {author} {\bibfnamefont {M.}~\bibnamefont {Gorchtein}}, \ and\ \bibinfo
  {author} {\bibfnamefont {M.}~\bibnamefont {Vanderhaeghen}},\ }\href {\doibase
  10.1103/PhysRevC.103.024001} {\bibfield  {journal} {\bibinfo  {journal}
  {Physical Review C}\ }\textbf {\bibinfo {volume} {103}},\ \bibinfo {pages}
  {024001} (\bibinfo {year} {2021})}\BibitemShut {NoStop}%
\bibitem [{\citenamefont {Hernandez}\ \emph {et~al.}(2019)\citenamefont
  {Hernandez}, \citenamefont {Ji}, \citenamefont {Bacca},\ and\ \citenamefont
  {Barnea}}]{hernandez2019probing}%
  \BibitemOpen
  \bibfield  {author} {\bibinfo {author} {\bibfnamefont {O.~J.}\ \bibnamefont
  {Hernandez}}, \bibinfo {author} {\bibfnamefont {C.}~\bibnamefont {Ji}},
  \bibinfo {author} {\bibfnamefont {S.}~\bibnamefont {Bacca}}, \ and\ \bibinfo
  {author} {\bibfnamefont {N.}~\bibnamefont {Barnea}},\ }\href {\doibase
  10.1103/PhysRevC.100.064315} {\bibfield  {journal} {\bibinfo  {journal}
  {Physical Review C}\ }\textbf {\bibinfo {volume} {100}},\ \bibinfo {pages}
  {064315} (\bibinfo {year} {2019})}\BibitemShut {NoStop}%
\bibitem [{\citenamefont {Vautherin}\ and\ \citenamefont
  {Veneroni}(1969)}]{Vautherin1969PLB29:203--206}%
  \BibitemOpen
  \bibfield  {author} {\bibinfo {author} {\bibfnamefont {D.}~\bibnamefont
  {Vautherin}}\ and\ \bibinfo {author} {\bibfnamefont {M.}~\bibnamefont
  {Veneroni}},\ }\href {\doibase https://doi.org/10.1016/0370-2693(69)90140-3}
  {\bibfield  {journal} {\bibinfo  {journal} {Phys. Lett. B}\ }\textbf
  {\bibinfo {volume} {29}},\ \bibinfo {pages} {203} (\bibinfo {year}
  {1969})}\BibitemShut {NoStop}%
\bibitem [{\citenamefont {Vautherin}\ and\ \citenamefont
  {Brink}(1972)}]{Vautherin1972PRC5:626--647}%
  \BibitemOpen
  \bibfield  {author} {\bibinfo {author} {\bibfnamefont {D.}~\bibnamefont
  {Vautherin}}\ and\ \bibinfo {author} {\bibfnamefont {D.~M.}\ \bibnamefont
  {Brink}},\ }\href {\doibase 10.1103/PhysRevC.5.626} {\bibfield  {journal}
  {\bibinfo  {journal} {Phys. Rev. C}\ }\textbf {\bibinfo {volume} {5}},\
  \bibinfo {pages} {626} (\bibinfo {year} {1972})}\BibitemShut {NoStop}%
\bibitem [{\citenamefont {Bender}\ \emph {et~al.}(2003)\citenamefont {Bender},
  \citenamefont {Heenen},\ and\ \citenamefont
  {Reinhard}}]{Bender2003RMP75:121--180}%
  \BibitemOpen
  \bibfield  {author} {\bibinfo {author} {\bibfnamefont {M.}~\bibnamefont
  {Bender}}, \bibinfo {author} {\bibfnamefont {P.-H.}\ \bibnamefont {Heenen}},
  \ and\ \bibinfo {author} {\bibfnamefont {P.-G.}\ \bibnamefont {Reinhard}},\
  }\href {\doibase 10.1103/RevModPhys.75.121} {\bibfield  {journal} {\bibinfo
  {journal} {Rev. Mod. Phys.}\ }\textbf {\bibinfo {volume} {75}},\ \bibinfo
  {pages} {121} (\bibinfo {year} {2003})}\BibitemShut {NoStop}%
\end{thebibliography}%

\end{document}